\documentclass[a4paper]{jpconf}
\usepackage{graphicx}
\usepackage{amssymb}
\usepackage{amsmath}

\usepackage{cite}

\begin{document}
\title{Extended Supersymmetries and $2+1$ Dimensional Supersymmetric Chern Simons Theories\footnote{Prepared for the Conference Integrable Systems and Quantum Symmetries
June 2013
Prague}}

\author{V K Oikonomou}

\address{Max Planck Institute for Mathematics in the Sciences\\
Inselstrasse 22, 04103 Leipzig, Germany}

\ead{voiko@physics.auth.gr}

\begin{abstract}
We study $N=2$ supersymmetric Chern-Simons Higgs models in $(2+1)$-dimensions. As we will demonstrate, an extended supersymmetric quantum mechanics algebras underlies the fermionic zero modes quantum system and the zero modes corresponding to bosonic fluctuations. These two algebras, in turn, combine to give an $N=4$ extended 1-dimensional supersymmetric algebra with central charge
\end{abstract}

Gauge field theories in $(2+1)$-dimensions \cite{dunnebook,lee} are particularly interesting, since the Chern-Simons \cite{lee1,lee1a,lee1b} term can be consistently incorporated in the theoretical framework. This term drastically alters the long distance behavior of the theory and modifies the allowed solitonic solutions. Specifically, the topologically stable solitonic solutions have electric charge \cite{lee3,lee3a,lee3b}. For some Higgs potentials, the minimum energy vortex solutions satisfy the Bogomol'nyi equations \cite{dunnebook,lee19,lee7,lee16}. These self-dual systems have static multisoliton solutions which are localized in different points in space, or superimposed at one point.

\noindent In reference to supersymmetric extensions of $(2+1)$-dimensional gauged Chern-Simons models, the fermionic zero modes represent the degeneracy of the solitonic states, while the zero modes of the bosonic fluctuations determine the collective coordinates which describe the solitons and their small velocity kinematics. As is well known \cite{lee}, in $N=2$ spacetime supersymmetric models, the fermionic zero modes are directly related to the zero modes of bosonic fluctuations. 

\noindent In this presentation (which is based on \cite{oikonomou}) we demonstrate that the zero modes of fermionic and bosonic fluctuations of Abelian gauge models having a global $N=2$ spacetime supersymmetry in $(2+1)$-dimensions, can separately constitute two $N=2$, $d=1$ supersymmetric quantum algebras \cite{witten,susyqm,susyqm1,susyqm2,susyqm3,susyqm4,susyqm5,thaller,susynew1,susynew2,susynew3,susynew4,susynew5,susynew6,susynew7,susynew8,susynew9}, with the zero modes being the corresponding quantum Hilbert space vectors. Due to the $N=2$ global spacetime supersymmetry of the theory, the two one-dimensional $N=2$ supersymmetric quantum algebras combine to form a centrally extended $N=4$, $d=1$ supersymmetric quantum algebra. Consequently, the quantum system of fermionic zero modes and of bosonic fluctuations zero modes, constitute an quantum system with centrally extended supersymmetry. This provides us with important information about the bosonic fluctuations subsystem and particularly for the quantum theory of the vortices. This is because the bosonic fluctuations correspond to collective coordinates describing the vortices positions and slow velocity kinematics. 

\noindent We shall present the models in the limiting case $\kappa =0$, with $\kappa$ the coupling of the Chern-Simons term. This case, corresponds to the supersymmetric extension of the Landau-Ginzburg model. Therefore, the results could provide us with interesting information, in reference to the quantum Hall effect and to the high-temperature superconductivity \cite{oh,oh1,oh2}.

The model that we will extensively 
use in this presentation, following reference \cite{lee,oikonomou}, are described by the following $N=2$ spacetime supersymmetric model Lagrangian,
\begin{align}\label{n2lag}
&\mathcal{L}=-\frac{1}{4}F_{\mu \nu}^{\mu \nu}+\frac{1}{4}\kappa \epsilon^{\mu \nu \lambda}F_{\mu \nu}A_{\lambda}-\lvert D_{\mu}\phi \lvert^{2}-\frac{1}{2}(\partial_{\mu}N)^2
\\ \notag & -\frac{1}{2}(e\lvert\phi\lvert^2+\kappa N-eu^2)^2-e^2N^2\lvert\phi\lvert^2
\\ \notag & i\bar{\psi}\gamma^{\mu}D_{\mu}\psi+i\bar{\chi}\gamma^{\mu}\partial_{\mu}\chi+\kappa\bar{\chi}\chi
+i\sqrt{2}e(\bar{\psi}\chi\phi-\bar{\chi}\psi\phi^*)+e N\bar{\psi}\psi
\end{align}
where $D_{\mu}=\partial_{\mu}-ieA_{\mu}$ and the gamma matrices satisfy the relation $\gamma^{\mu}\gamma^{\nu}=-\eta^{\mu\nu}-i\epsilon^{\mu \nu \lambda}\gamma_{\lambda}$. The fields $\psi ,\chi$ are two component spinors, with $\psi$ being a Weyl charged fermionic field and $\chi$ a neutral complex Weyl two component spinor. We turn our focus on the fermionic part of the $N=2$ Lagrangian in order to find the fermionic zero modes, in the background of self dual vortices. The fermionic equations of motion corresponding to Lagrangian (\ref{n2lag}), are:
\begin{align}\label{eqnmotion}
& \gamma^i D_i \psi+ie(\gamma^0 A^0-N)\psi-\sqrt{2}e\phi \chi=0
\\ \notag & \gamma^i\partial_i \chi -i\kappa \chi+\sqrt{2}e\phi^*\psi=0
\end{align}
Making the following conventions:
\begin{equation}\label{conventions}
\gamma^0=\sigma_3,{\,}{\,}\gamma^1=i\sigma_2,{\,}{\,}\gamma^2=i\sigma_1
\end{equation}
and by setting:
\begin{equation}\label{setting}
\psi=\left (\begin{array}{c}
        \psi_{\uparrow}          \\
    \psi_{\downarrow} \\
\end{array}\right ),{\,}{\,}{\,}\chi=\left (\begin{array}{c}
        \chi_{\uparrow}          \\
    \chi_{\downarrow} \\
\end{array}\right )
\end{equation}
and moreover assuming positive values of the flux, the equations of motion for the fermions can be cast in the following form:
\begin{align}\label{fermionsequationsofmotions}
&(D_1+iD_2)\psi_{\downarrow}-\sqrt{2}e\phi \chi_{\uparrow}=0
\\ \notag & (\partial_1-i\partial_2)\chi_{\uparrow }+i\kappa \chi_{\downarrow }-\sqrt{2}\phi^*\psi_{\downarrow }=0
\\ \notag & (D_1-iD_2)\psi_{\uparrow }+2i e A^0\psi_{\downarrow }+\sqrt{2}e\phi \chi_{\downarrow }=0
\\ \notag & (\partial_1+i\partial_2)\chi_{\downarrow }-i\kappa \chi_{\uparrow }+\sqrt{2}e\phi^*\psi_{\uparrow }=0
\end{align}
Now,  in the $\kappa =0$ case, the fermionic equations of motion (\ref{fermionsequationsofmotions}) become,
\begin{align}\label{fer1}
&(D_1+iD_2)\psi_{\downarrow}-\sqrt{2}e\phi\chi_{\uparrow}=0
\\ \notag & (\partial_1-i\partial_2)\chi_{\uparrow}-\sqrt{2}e\phi^*\psi_{\downarrow}=0
\\ \notag & (D_1-iD_2)\psi_{\uparrow}+\sqrt{2}e\phi\chi_{\downarrow}=0
\\ \notag & (\partial_1+i\partial_2)\chi_{\downarrow}+\sqrt{2}e\phi^*\psi_{\uparrow}=0
\end{align}

The last two equations of relation (\ref{fer1}) have no localized fermionic solutions, apart from some trivial solutions \cite{lee}. Conversely, the first two equations of relation (\ref{fer1}), have $2n$ solutions, with $n$ the vorticity number \cite{lee}. We can form an operator $\mathcal{D}_{LG}$, corresponding to the first two equations of (\ref{fer1}),
\begin{equation}\label{susyqmrn567m}
\mathcal{D}_{LG}=\left(%
\begin{array}{cc}
 D_1+iD_2 & -\sqrt{2}e\phi
 \\ -\sqrt{2}\phi^* & \partial_1-i\partial_2\\
\end{array}%
\right)
\end{equation}
which acts on the vector:
\begin{equation}\label{ait34e1}
|\Psi_{LG}\rangle =\left(%
\begin{array}{c}
  \psi_{\downarrow} \\
  \chi_{\uparrow} \\
\end{array}%
\right).
\end{equation}
Therefore, the first two equations of (\ref{fer1}) can be written as:
\begin{equation}\label{transf}
\mathcal{D}_{LG}|\Psi_{LG}\rangle=0
\end{equation}
The solutions of the above equation are the zero modes of the operator $\mathcal{D}_{LG}$, but since the first two equations of (\ref{fer1}) have $2n$ normalized solutions, we may conclude that:
\begin{equation}\label{dimeker}
\mathrm{dim}{\,}\mathrm{ker}\mathcal{D}_{LG}=2n
\end{equation}
Furthermore, the adjoint of the operator $\mathcal{D}_{LG}$, namely $\mathcal{D}_{LG}^{\dag}$, is equal to:
\begin{equation}\label{eqndag}
\mathcal{D}_{LG}^{\dag}=\left(%
\begin{array}{cc}
 D_1-iD_2 & \sqrt{2}e\phi
 \\ \sqrt{2}\phi^* & \partial_1+i\partial_2\\
\end{array}%
\right)
\end{equation}
and acts on the vector:
\begin{equation}\label{ait3hgjhgj4e1}
|\Psi_{LG}'\rangle =\left(%
\begin{array}{c}
  \psi_{\uparrow} \\
  \chi_{\downarrow} \\
\end{array}%
\right).
\end{equation}


The zero modes of the operator $\mathcal{D}_{LG}^{\dag}$, are identical to the solutions of the last two equations of (\ref{fer1}), with the obvious replacement $e\rightarrow-e$. Since the corresponding pair of equations has no normalized solutions, the kernel of the operator $\mathcal{D}_{LG}^{\dag}$ is null:
\begin{equation}\label{dimeke1r11}
\mathrm{dim}{\,}\mathrm{ker}\mathcal{D}_{LG}^{\dag}=0
\end{equation}
The fermionic system in the self dual vortices background, has an unbroken $N=2$, $d=1$ supersymmetry. The supercharges of this $N=2$, $d=1$ SUSY algebra are equal to:
\begin{equation}\label{s7}
\mathcal{Q}_{LG}=\bigg{(}\begin{array}{ccc}
  0 & \mathcal{D}_{LG} \\
  0 & 0  \\
\end{array}\bigg{)},{\,}{\,}{\,}\mathcal{Q}^{\dag}_{LG}=\bigg{(}\begin{array}{ccc}
  0 & 0 \\
  \mathcal{D}_{LG}^{\dag} & 0  \\
\end{array}\bigg{)}
\end{equation}
while the quantum Hamiltonian of the quantum system is:
\begin{equation}\label{s11}
\mathcal{H}_{LG}=\bigg{(}\begin{array}{ccc}
 \mathcal{D}_{LG}\mathcal{D}_{LG}^{\dag} & 0 \\
  0 & \mathcal{D}_{LG}^{\dag}\mathcal{D}_{LG}  \\
\end{array}\bigg{)}
\end{equation}
These elements of the algebra, satisfy the $d=1$ SUSY algebra:
\begin{equation}\label{relationsforsusy}
\{\mathcal{Q}_{LG},\mathcal{Q}^{\dag}_{LG}\}=\mathcal{H}_{LG}{\,}{\,},\mathcal{Q}_{LG}^2=0,{\,}{\,}{\mathcal{Q}_{LG}^{\dag}}^2=0
\end{equation}
The Hilbert space of the supersymmetric quantum mechanical system, $\mathcal{H}_{LG}$ is $Z_2$ graded by the operator $\mathcal{W}$, the Witten parity, an involution operator which commutes with the total Hamiltonian,
\begin{equation}\label{s45}
[\mathcal{W},\mathcal{H}_{LG}]=0
\end{equation}
and  anti-commutes with the supercharges,
\begin{equation}\label{s5}
\{\mathcal{W},\mathcal{Q}_{LG}\}=\{\mathcal{W},\mathcal{Q}_{LG}^{\dag}\}=0
\end{equation}
In addition, the operator $\mathcal{W}$ satisfies the following identity,
\begin{equation}\label{s6}
\mathcal{W}^{2}=1
\end{equation}
The Witten parity $\mathcal{W}$, spans the total Hilbert space into $Z_2$ equivalent subspaces. Therefore, the total Hilbert space of the quantum system can be written as:
\begin{equation}\label{fgjhil}
\mathcal{H}=\mathcal{H}^+\oplus \mathcal{H}^-
\end{equation}
with the vectors that belong to the two subspaces $\mathcal{H}^{\pm}$, classified according to their Witten parity, to even and odd parity states, that is:
\begin{equation}\label{shoes}
\mathcal{H}^{\pm}=\mathcal{P}^{\pm}\mathcal{H}=\{|\psi\rangle :
\mathcal{W}|\psi\rangle=\pm |\psi\rangle \}
\end{equation}
Moreover, the Hamiltonians of the $Z_2$ graded spaces are equal to:
\begin{equation}\label{h1}
{\mathcal{H}}_{+}=\mathcal{D}_{LG}{\,}\mathcal{D}_{LG}^{\dag},{\,}{\,}{\,}{\,}{\,}{\,}{\,}{\mathcal{H}}_{-}=\mathcal{D}_{LG}^{\dag}{\,}\mathcal{D}_{LG}
\end{equation}
The operator $\mathcal{W}$ can be represented in the following matrix form:
\begin{equation}\label{wittndrf}
\mathcal{W}=\bigg{(}\begin{array}{ccc}
  1 & 0 \\
  0 & -1  \\
\end{array}\bigg{)}
\end{equation}
The eigenstates of the operator $\mathcal{P}^{\pm}$, namely $|\psi^{\pm}\rangle$, satisfy the following relation:
\begin{equation}\label{fd1}
P^{\pm}|\psi^{\pm}\rangle =\pm |\psi^{\pm}\rangle
\end{equation}
We call them positive and negative parity eigenstates, with ``parity'' referring to the $P^{\pm}$ operator, which is nothing else but the Witten parity operator. Using the representation (\ref{wittndrf}) for the Witten parity operator,
the parity eigenstates can represented by the vectors,
\begin{equation}\label{phi5}
|\psi^{+}\rangle =\left(%
\begin{array}{c}
  |\phi^{+}\rangle \\
  0 \\
\end{array}%
\right),{\,}{\,}{\,}
|\psi^{-}\rangle =\left(%
\begin{array}{c}
  0 \\
  |\phi^{-}\rangle \\
\end{array}%
\right)
\end{equation}
with $|\phi^{\pm}\rangle$ $\epsilon$ $\mathcal{H}^{\pm}$. We can write the fermionic states of the system in terms of the SUSY quantum algebra:
\begin{equation}\label{fdgdfgh}
|\Psi_{LG}\rangle =|\phi^{-}\rangle=\left(%
\begin{array}{c}
  \psi_{\downarrow} \\
  \chi_{\uparrow} \\
\end{array}%
\right),{\,}{\,}{\,}|\Psi_{LG}'\rangle =|\phi^{+}\rangle=\left(%
\begin{array}{c}
  \psi_{\uparrow} \\
  \chi_{\downarrow} \\
\end{array}%
\right)
\end{equation}
Thereby, the even and odd parity SUSY quantum states are:
\begin{equation}\label{phi5}
|\psi^{+}\rangle =\left(%
\begin{array}{c}
  |\Psi_{LG}'\rangle \\
  0 \\
\end{array}%
\right),{\,}{\,}{\,}
|\psi^{-}\rangle =\left(%
\begin{array}{c}
  0 \\
  |\Psi_{LG}\rangle \\
\end{array}%
\right)
\end{equation}
upon which, the Hamiltonian and the supercharges act. Supersymmetry is considered unbroken if the Witten index is a non-zero integer. The Witten index for Fredholm operators is defined to be:
\begin{equation}\label{phil}
\Delta =n_{-}-n_{+}
\end{equation}
with $n_{\pm}$ the number of zero
modes of ${\mathcal{H}}_{\pm}$ in the subspace $\mathcal{H}^{\pm}$, with the constraint that these are finitely many.

\noindent The case for which the Witten index is zero and also
if $n_{+}=n_{-}=0$, then supersymmetry is broken. Conversely, if $n_{+}=
n_{-}\neq 0$, the system has still an unbroken supersymmetry.

\noindent The Witten index is directly related to the Fredholm index of the operator $\mathcal{D}_{LG}$, as can be seen in the following equations:
\begin{align}\label{ker1}
&\Delta=\mathrm{dim}{\,}\mathrm{ker}
{\mathcal{H}}_{-}-\mathrm{dim}{\,}\mathrm{ker} {\mathcal{H}}_{+}=
\mathrm{dim}{\,}\mathrm{ker}\mathcal{D}_{LG}^{\dag}\mathcal{D}_{LG}-\mathrm{dim}{\,}\mathrm{ker}\mathcal{D}_{LG}\mathcal{D}_{LG}^{\dag}=
\\ \notag & \mathrm{ind} \mathcal{D}_{LG} = \mathrm{dim}{\,}\mathrm{ker}
\mathcal{D}_{LG}-\mathrm{dim}{\,}\mathrm{ker} \mathcal{D}_{LG}^{\dag}
\end{align}
Recalling the results of equations (\ref{dimeker}) and (\ref{dimeke1r11}), the Witten index is equal to:
\begin{equation}\label{fnwitten}
\Delta =-2n
\end{equation}
Therefore, the fermionic system in the self-dual Landau-Ginzburg vortices background with $N=2$ spacetime supersymmetry, has an unbroken $N=2$, $d=1$ supersymmetry. This could be owing to the fact that the initial system has an unbroken $N=2$ spacetime supersymmetry, so the Hilbert space of the zero modes states retains a $N=2$, $d=1$ supersymmetric quantum algebra. However this is not true. Supersymmetric algebras with equal number of supercharges, but in different dimensions are a different concept. For example, the spacetime supersymmetric algebra is four dimensional while the SUSY quantum mechanics algebra is one dimensional. There is an obvious correlation between the two supersymmetries, since extended (with $N = 4, 6...$) supersymmetric quantum mechanics models can describe
the dimensional reduction to one (temporal) dimension of N = 2 and N = 1
Super-Yang Mills models \cite{ivanov1,ivanov2}. However, the $N = 2$, $d=1$ SUSY quantum mechanics supercharges
do not generate spacetime supersymmetry. In addition, the supersymmetry in
supersymmetric quantum mechanics doesn't relates fermions and bosons, but SUSY QM supercharges classify the Hilbert space of quantum states according to the group $Z_2$. Moreover,
the supercharges generate transformations between the Witten parity eigenstates, which are two orthogonal eigenstates of the quantum Hamiltonian. Finally, the supersymmetric quantum mechanics algebra is not a spacetime supersymmetry, but a quantum algebra that some of the representations of the super-Poincare algebra obey.

The bosonic zero modes equations for $\kappa =0$ and $A^0=N=0$ are cast in the following form:
\begin{align}\label{selfdualfluctuat1fd}
&(D_1+iD_2)\delta \phi-ie\phi (\delta A_1+i \delta A_2)=0
\\ \notag &(\partial_1-\partial_2)(\delta A_1+ i\delta A_2)+2ie\phi^*\delta \phi =0
\end{align}
Note that the above equations become identical to equations (\ref{fer1}), if we substitute:
\begin{equation}\label{biossubs}
\psi_{\downarrow}=\delta \phi,{\,}{\,}{\,}\chi_{\uparrow}=\frac{i}{\sqrt{2}}(\delta A_1+i\delta A_2)
\end{equation}
Therefore, we conclude that the number of the zero modes corresponding to equation (\ref{selfdualfluctuat1fd}), is equal to $2n$. As in the fermionic case, an unbroken $N=2$ SUSY quantum mechanical algebra underlies the bosonic system as well. Indeed, equations (\ref{selfdualfluctuat1fd}) can be written in the following form:
\begin{equation}\label{transfrey}
\mathcal{D}_{LG}'|\Phi_{LG}\rangle=0
\end{equation}
with the operator $\mathcal{D}_{LG}'$ being equal to:
\begin{equation}\label{susyqmrrtyurn567m}
\mathcal{D}_{LG}'=\left(%
\begin{array}{cc}
 D_1+iD_2 & -\sqrt{2}e\phi
 \\ -\sqrt{2}\phi^* & \partial_1-i\partial_2\\
\end{array}%
\right)
\end{equation}
and is considered to act on the vector:
\begin{equation}\label{ait3urtu4e1}
|\Phi_{LG}\rangle =\left(%
\begin{array}{c}
  \delta \phi \\
  \frac{i}{\sqrt{2}}(\delta A_1+i\delta A_2) \\
\end{array}%
\right).
\end{equation}
Hence:
\begin{equation}\label{dimektriier}
\mathrm{dim}{\,}\mathrm{ker}\mathcal{D}_{LG}'=2n
\end{equation}
As in the fermionic case,
\begin{equation}\label{dimekegkjl1r11}
\mathrm{dim}{\,}\mathrm{ker}{\mathcal{D}_{LG}'}^{\dag}=0
\end{equation}
The operators ${\mathcal{D}_{LG}'}$ and ${\mathcal{D}_{LG}'}^{\dag}$ are Fredholm, and therefore any which operator constructed from these, is also Fredholm. The supercharges and the Hamiltonian, that constitute the $N=2$, $d=1$ algebra in the bosonic case, are:
\begin{equation}\label{sgggggg7}
\mathcal{Q}_{LG}'=\bigg{(}\begin{array}{ccc}
  0 & \mathcal{D}_{LG}' \\
  0 & 0  \\
\end{array}\bigg{)},{\,}{\,}{\,}{{\mathcal{Q}'}^{\dag}}_{LG}=\bigg{(}\begin{array}{ccc}
  0 & 0 \\
  {\mathcal{D}'}_{LG}^{\dag} & 0  \\
\end{array}\bigg{)},{\,}{\,}{\,}\mathcal{H}_{LG}'=\bigg{(}\begin{array}{ccc}
 \mathcal{D}_{LG}\mathcal{D}_{LG}^{\dag} & 0 \\
  0 & \mathcal{D}_{LG}^{\dag}\mathcal{D}_{LG}  \\
\end{array}\bigg{)}
\end{equation}

\noindent The elements of the algebra, satisfy the $d=1$ SUSY algebra:
\begin{equation}\label{relationsforsusy}
\{\mathcal{Q}_{LG}',{\mathcal{Q}'}^{\dag}_{LG}\}=\mathcal{H}_{LG}'{\,}{\,},\mathcal{Q}_{LG}^2=0,{\,}{\,}{\mathcal{Q}_{LG}^{\dag}}^2=0
\end{equation}
Supersymmetry is unbroken, since the corresponding Witten index $\Delta '$ is a non-zero integer, in this case too. Indeed:
\begin{equation}\label{wittindexfgo}
\Delta '= -2n
\end{equation}
Hence, we can see that an unbroken $N=2$ SUSY quantum mechanics algebra underlies the bosonic fluctuations in the self dual Landau-Ginzburg vortex background. This algebra is identical to the fermionic SUSY quantum mechanics algebra. It is reasonable to ask if there is a more rich symmetry structure in the fermion-boson system, apart from these two $N=2$ algebras. The answer is yes. Indeed, we can see this if we calculate the following commutation and anti-commutation relations:
\begin{align}\label{commutatorsanticomm}
&\{{{\mathcal{Q}'}_{LG}},{{\mathcal{Q}'}_{LG}}^{\dag}\}=2\mathcal{H},{\,}\{{{\mathcal{Q}}_{LG}},{{\mathcal{Q}}_{LG}}^{\dag}\}=2\mathcal{H},{\,}\{{{\mathcal{Q}}_{LG}},{{\mathcal{Q}}_{LG}}\}=0,{\,}\{{{\mathcal{Q}'}_{LG}},{{\mathcal{Q}'}_{LG}}\}=0,{\,}{\,}\\
\notag & \{{{\mathcal{Q}}_{LG}},{{\mathcal{Q}'}_{LG}}^{\dag}\}=\mathcal{Z},{\,}\{{{\mathcal{Q}'}_{LG}},{{\mathcal{Q}}_{LG}}^{\dag}\}=\mathcal{Z},{\,}\\ \notag
&\{{{\mathcal{Q}'}_{LG}}^{\dag},{{\mathcal{Q}'}_{LG}}^{\dag}\}=0,\{{{\mathcal{Q}}_{LG}}^{\dag},{{\mathcal{Q}}_{LG}}^{\dag}\}=0,{\,}\{{{\mathcal{Q}}_{LG}}^{\dag},{{\mathcal{Q}'}_{LG}}^{\dag}\}=0,{\,}\{{{\mathcal{Q}}_{LG}},{{\mathcal{Q}'}_{LG}}\}=0{\,}\\
\notag
&[{{\mathcal{Q}'}_{LG}},{{\mathcal{Q}}_{LG}}]=0,[{{\mathcal{Q}}_{LG}}^{\dag},{{\mathcal{Q}'}_{LG}}^{\dag}]=0,{\,}[{{\mathcal{Q}'}_{LG}},{{\mathcal{Q}'}_{LG}}]=0{\,}[{{\mathcal{Q}'}_{LG}}^{\dag},{{\mathcal{Q}'}_{LG}}^{\dag}]=0,{\,}\\
\notag &
[{\mathcal{H}'}_{LG},{{\mathcal{Q}'}_{LG}}]=0,{\,}[{\mathcal{H}'}_{LG},{{\mathcal{Q}'}_{LG}}^{\dag}]=0,{\,}[\mathcal{H}_{LG},{{\mathcal{Q}}_{LG}}^{\dag}]=0,{\,}[\mathcal{H}_{LG},{{\mathcal{Q}}_{LG}}]=0,{\,}
\end{align}
with $\mathcal{Z}$:
\begin{equation}\label{zcentralcharge}
\mathcal{Z}=2\mathcal{H}_{LG}=2{\mathcal{H}'}_{LG}
\end{equation}
The operator $\mathcal{Z}$ commutes with all the elements of the two algebras, namely the
supercharges ${{\mathcal{Q}}_{LG}},{{\mathcal{Q}'}_{LG}}$, their conjugates ${{\mathcal{Q}}_{LG}}^{\dag},{{\mathcal{Q}'}_{LG}}^{\dag}$ and
finally the Hamiltonians, $\mathcal{H}=\mathcal{H}_{LG}={\mathcal{H}'}_{LG}$.
The relations
(\ref{commutatorsanticomm}) describe an $N=4$ supersymmetric quantum mechanics
algebra with central charge $\mathcal{Z}$, which is the Hamiltonian of each $N=2$ subsystem. In general, for an $N=4$ algebra we have:
\begin{align}\label{n4algbe}
&\{Q_i,Q_j^{\dag}\}=2\delta_i^jH+Z_{ij},{\,}{\,}i=1,2 \\ \notag &
\{Q_i,Q_j\}=0,{\,}{\,}\{Q_i^{\dag},Q_j^{\dag}\}=0
\end{align}
The algebra (\ref{commutatorsanticomm}) possesses two central charges which
are $Z_{12}=Z_{21}=\mathcal{Z}$.

Obviously, such a algebraic structure is interesting, since the fermions and the fluctuations of the bosonic fields, that are connected with an $N=2$ spacetime supersymmetric algebra, in $(2+1)$-dimensions, form the Hilbert space of an $N=4$ $d=1$ supersymmetric quantum mechanics algebra.


\end{document}